\documentclass[twocolumn]{elsart}
\usepackage[dvips]{graphicx}
\begin{document}
\begin{frontmatter}
\title{Measurements of spatial resolution of ATLAS Pixel detectors}
\author{Tommaso Lari\thanksref{ATLAS}}
\address{Dipartimento di Fisica, Universit\`a di Milano and INFN, Sezione di 
Milano, via Celoria 16, I-20133 Milano, Italy}
\thanks[ATLAS]{On Behalf of the ATLAS collaboration \cite{ATLAS}}
\begin{abstract}
Standard as well as irradiated silicon pixel detectors
developed for the ATLAS experiment were tested in a beam.
Digital and analog resolutions were  
determined comparing the positions
measured by a microstrip telescope and by the pixel detector. 
Digital resolutions of $10  
\; \mu \mathrm{m}$ and analog resolutions of 
$6 \; \mu \mathrm{m}$ (before irradiation) and $10 
\; \mu \mathrm{m}$ (after irradiation) 
are obtained without subtracting the error on the position determined by the 
telescope.
\end{abstract}
\date{8 September 2000} 
\end{frontmatter}

\section{Introduction}

The ATLAS Pixel sensors \cite{ATLAS,Ala99}
consist of $n^+$ implants on a high resistivity 
{\em n}-bulk substrate which turns {\em p}-type after irradiation. 
The pixel dimensions are $50 \times 400 \; \mu \mathrm{m}^2$. 
Results on three different sensor designs 
are presented here: one with  
{\em p-stop} isolation between pixels \cite{Bat89}, one with   
{\em p-spray} isolation \cite{Ric96} and one with {\em p-spray} isolation
and the pixels surrounded by a 
floating $n^+$ implantation. They will be referred as {\em p-stop}, 
new {\em p-spray} and old  {\em p-spray} designs. 
The second is the nearest to final design for ATLAS. 

The pulse height is obtained by the Time Over Threshold technique \cite{ATLAS}.
Typical thresholds were around 
3000 electrons with a  dispersion of 170 electrons rms. 
The noise was typically 150 electrons rms. 

Several single chip assemblies  
were characterised extensively in test beam experiments \cite{Rag00}. 
Only resolution measurements are discussed here.
A telescope consisting of 4 planes of silicon microstrips was used 
to reconstruct the tracks of the incident beam particles.
The setup allowed the 
selection of the angle $\alpha$ between the normal to the 
pixel sensor plane and the beam 
direction.  

Some 
sensors were irradiated to fluences comparable to those expected for LHC.
Resolution measurements are presented for a {\em p-spray}
device (old design) irradiated to $10^{15} \; \mathrm{n_{eq}/cm^2}$.    
Not irradiated sensors were operated at 150 V bias voltage achieving 
full depletion (280 $\mu$m).   
The irradiated sensor  
was operated at 600 V corresponding to a depletion depth of 190 $\mu$m
\cite{Rag00}. 
The average charge collected at the pixel centre for a perpendicular track
was thus reduced from 28000 to 15000 electrons. 

\section{Results}

In what follows, $x$ is the direction along 
the short ($50 \; \mu$m) side of the pixel, 
$y$ the $400 \; \mu$m direction. Beam direction, 
the normal to the detector and the $x$ axis 
lie in the same plane. 

The hit position can be reconstructed using only the position of fired pixels
({\em digital algorithm}). For cluster sizes  
greater than one it is
possible to use also the {\em analog algorithm} which takes into account  
the correlation between the track position and the variable 
$\eta= Q_l/(Q_l+Q_f)$ where $Q_f$ and $Q_l$ are the charges collected 
by the first and last pixels in the cluster. 
The charge collected by other (central) pixels  
is not correlated to track position and so it is not considered. 
More details and a comparison with alternative algorithms can
be found in  \cite{TurCla}. 
The degradation of resolution due to Landau fluctuations is 
decreased by setting to a maximum value 
$Q_{\mathrm{cut}}$ charges exceeding that limit.

\begin{figure}[!h]
\begin{center}
\includegraphics[width=7cm]{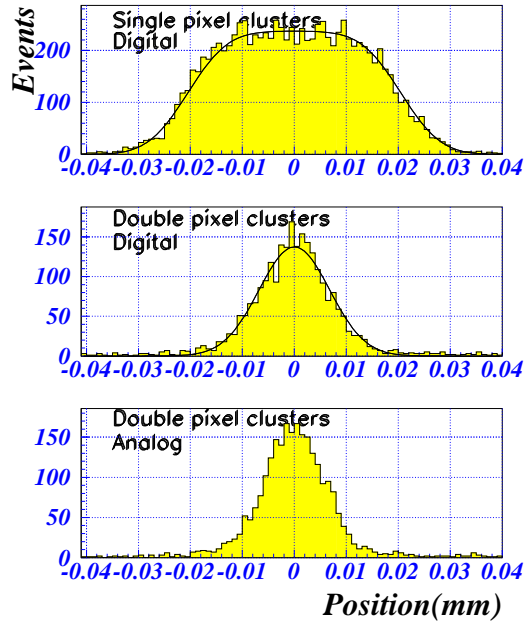}
\end{center}
\caption{Residuals \label{fig1} 
between position measured by the telescope and by pixel 
detector for the old {\em p-spray} 
device at $0^0$.}
\end{figure}

Fig. \ref{fig1} shows the $x$ residuals between position measured by 
the pixel detector and by the telescope  for the old {\em p-spray} 
device at $0^0$. At this angle 98\% of the clusters have one or two 
pixels. The few events with higher multiplicity are due to $delta$-ray
emission. At any given angle, analogously, 
there are always two dominant multiplicities. 
Residuals are plotted separately for each multiplicity.
  
\begin{figure}[!ht]
\begin{center}
\includegraphics[width=7cm,height=9cm]{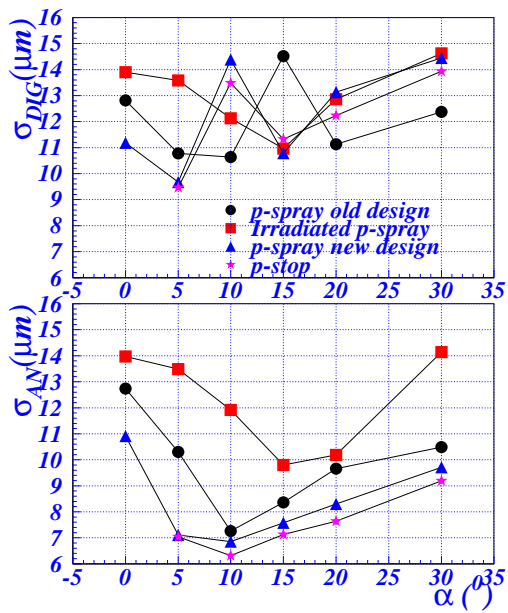}
\end{center}
\label{res}
\caption{Digital (above) and analog (below) residuals as a function 
of the angle of incidence of the beam, before subtraction of telescope 
resolution.}
\end{figure}

Digital residual distributions 
are fitted with a square convoluted with a gaussian. The gaussian width 
is taken to be the telescope extrapolation error. Values 
of telescope resolution between 5.4 and 6.4
$\mu$m are found for $\alpha \le 10^0$. 
The square length is 
the range of track positions corresponding to that multiplicity. For old 
{\em p-spray} devices at $0^0$ this range is a 
$41 \; \mu$m interval centred on the centre of the pixel for single pixel 
clusters, a  $9 \; \mu$m interval centred on the border between the pixels 
for double pixel clusters. 
The best digital resolution (Fig. \ref{res}) 
is obtained at angles for which the two dominant multiplicities  
are equally populated ($5^0$ for not irradiated and $15^0$ for 
irradiated devices) while the worst resolution occurs when one of them 
is much more populated (as in Fig. \ref{fig1}).

At $5^0$ for 
not irradiated and $15^0$ for irradiated devices single and double 
pixel cluster occur with the same frequency and resolution is good.  
At $10^0$ for {\em p-stop} devices mainly double pixel clusters 
occur and resolution is worse.    wermes@opala7.physik.uni-bonn.de

The best analog resolution (Fig. \ref{res}) is 
found at the angle at which mainly double pixel clusters 
occur. A resolution of $6.3 \; \mu$m is obtained at 
$10^0$ for the {\em p-stop}
device which has the best charge collection, while a resolution of 
$9.8 \; \mu$m at $15^0$ is obtained for the irradiated sensor.

After deconvolution of the telescope error the digital resolution 
is always in the $(25 \div 
50)/\sqrt{12} \; \mu \mathrm{m}$ interval, 
analog resolutions of 
$8 \div 12 \; \mu$m for the irradiated sensor and lower than  
$\sim 4 \; \mu$m for not-irradiated {\em p-stop} 
and new {\em p-spray} sensors at $10^0$ are found. 

\begin{figure}[!hb]
\begin{center}
\includegraphics[width=7cm]{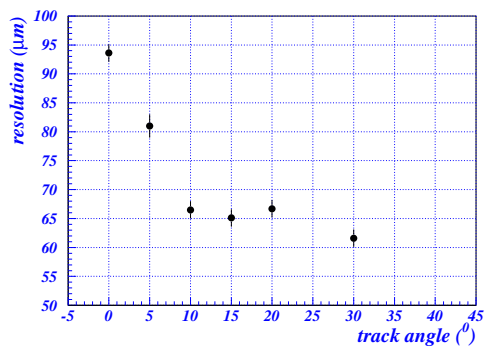}
\end{center}
\caption{Resolution in $y$ direction as a function of angle 
of incidence of the beam for the bricked design.
\label{bricked}} 
\end{figure}

In the $y$ direction the residuals present a 
flat distribution between -200 $\mu$m to +200 $\mu$m with a rms 
of 115 $\mu$m. 
A bricked structure (with 
pixels in adjacent rows displaced by 200 $\mu$m) was also tested with the 
aim of improving the y resolution by a factor of two    
for clusters with charge sharing in the $x$ direction. 
Fig. \ref{bricked} shows the $y$ resolution as a function of the angle. 
For $\alpha \ge 10^0$ all clusters have at least 
two pixels and the resolution reaches 65 $\mu$m. 

\section{Conclusions}
Resolution measurements have been done on 
standard as well as 
irradiated prototypes of ATLAS Pixel sensors.
In the short direction of the cell analog resolutions 
in the range 6-14 $\mu$m have been measured before 
subtraction of telescope resolution, estimated to be about 6 $\mu$m.
A bricked design has been shown to improve the resolution in the 
long direction of the cell.

\end{document}